\documentclass[conference]{IEEEtran}
\ifCLASSINFOpdf
\else
\fi
\usepackage{multirow}    % Use to create table with a column that spans multiple rows
\usepackage{pifont}      % Provides the ding symbol used for comments
\usepackage{color}       % Used to highlight comments
\usepackage{xspace}      % Intelligently adds space after a word via \xspace
\usepackage{flushend}     % balance the last page
\usepackage{graphicx}
\usepackage{caption}
\usepackage{subcaption}
\usepackage{url}
\usepackage{rotating}
\usepackage{paralist}
\usepackage{soul}
\usepackage[linesnumbered,ruled]{algorithm2e}
\makeatletter
\renewcommand{\@algocf@capt@plain}{above}% formerly {bottom}
\makeatother % for algorithmic

\usepackage{algorithmicx}  
\usepackage{algpseudocode}  
\usepackage{amsmath,amsthm} 
\algnewcommand{\LineComment}[1]{\State \(\triangleright\) #1}

\usepackage{tikz}
\usepackage{xcolor}

\usepackage[normalem]{ulem}

\newcommand{\circled}[1]{\tikz[baseline=(myanchor.base)] \node[circle,fill=.,inner sep=1pt] (myanchor) {\color{-.}\bfseries\footnotesize #1};}
\newcommand{\mypara}[1]{\vspace{2pt}\noindent\textbf{{#1: }}}
\newcommand{\eat}[1]{}  %% for quick commenting of a large trunk of texts
\newcommand{\name}{GhostKnight\xspace}

\begin{document}
\title{\name: Breaching Data Integrity via Speculative Execution}


\author{
\IEEEauthorblockN{Zhi Zhang$^{*,1}$, Yueqiang Cheng$^{*,2}$, and Surya Nepal$^1$}
\IEEEauthorblockA{The draft is subject to change\\
$^*$ Both authors contributed equally to this work \\ $^1$ Data61, CSIRO, Australia \\ $^2$ NIO \\}
}


%Telephone: (800) 555--1212, Fax: (888) 555--1212}


\maketitle

\begin{abstract}
%The primary contribution of this work is to show that Spectre-like attacks pose a threat not only to integrity,  but  also to confidentiality as well.
Existing speculative execution attacks are limited to breaching confidentiality of data beyond privilege boundary, the so-called spectre-type attacks. All of them utilize the changes in microarchitectural buffers made by the speculative execution to leak data. 
We show that the speculative execution can be abused to break data integrity.
We observe that the speculative execution not only leaves traces in the microarchitectural buffers but also induces side effects within DRAM, that is, the speculative execution can trigger an access to an illegitimate address in DRAM.
%In the case of existing speculative execution attacks (spectre-type exploits), an attacker is limited to . 
If the access to DRAM is frequent enough, then architectural changes (i.e., permanent bit flips in DRAM) will occur, which we term \name. 
%We demonstrate \name on Ubuntu Linux and it is able to induce the first bit flip with 5 minutes.  
%With the power of speculative execution, \name is also able to bypass the state-of-the-art software rowhammer defenses. 
With the power of of \name, an attacker is essentially able to cross different privilege boundaries and write exploitable bits to other privilege domains. 
In our future work, we will develop a \name-based exploit to cross a trusted execution environment, defeat a 1024-bit RSA exponentiation implementation and obtain a controllable signature.
\end{abstract}

\section{Introduction}\label{sec:intro}

In 2018, Kocher et al.~\cite{kocher2018spectre} uncovered a class of security vulnerabilities, the so-called ``spectre'', that embeds within the speculative execution of modern processors, 
extracting private information through a timing side-channel.

For instance, when the processor encounters a conditional branch, it depends on a value to determine the destination of the branch, but unfortunately the value is located in DRAM (Dynamic Random Access Memory) and the processor is supposed to wait for it to arrive. To maximize performance, the processor instead makes an educated guess about the destination and thus speculatively executes ahead. If the guess is correct, then a performance advantage is achieved. If the guess turns out to be wrong, then the processor discards the architectural changes (e.g., register values) caused by the speculative execution.
However, it leaves sensitive microarchitectural changes (e.g., data in cache), which are although invisible but can be leaked through known attack vectors such as side-channels~\cite{irazoqui2015s,liu2015last}.

\begin{figure}
\centering
\includegraphics[width=\columnwidth]{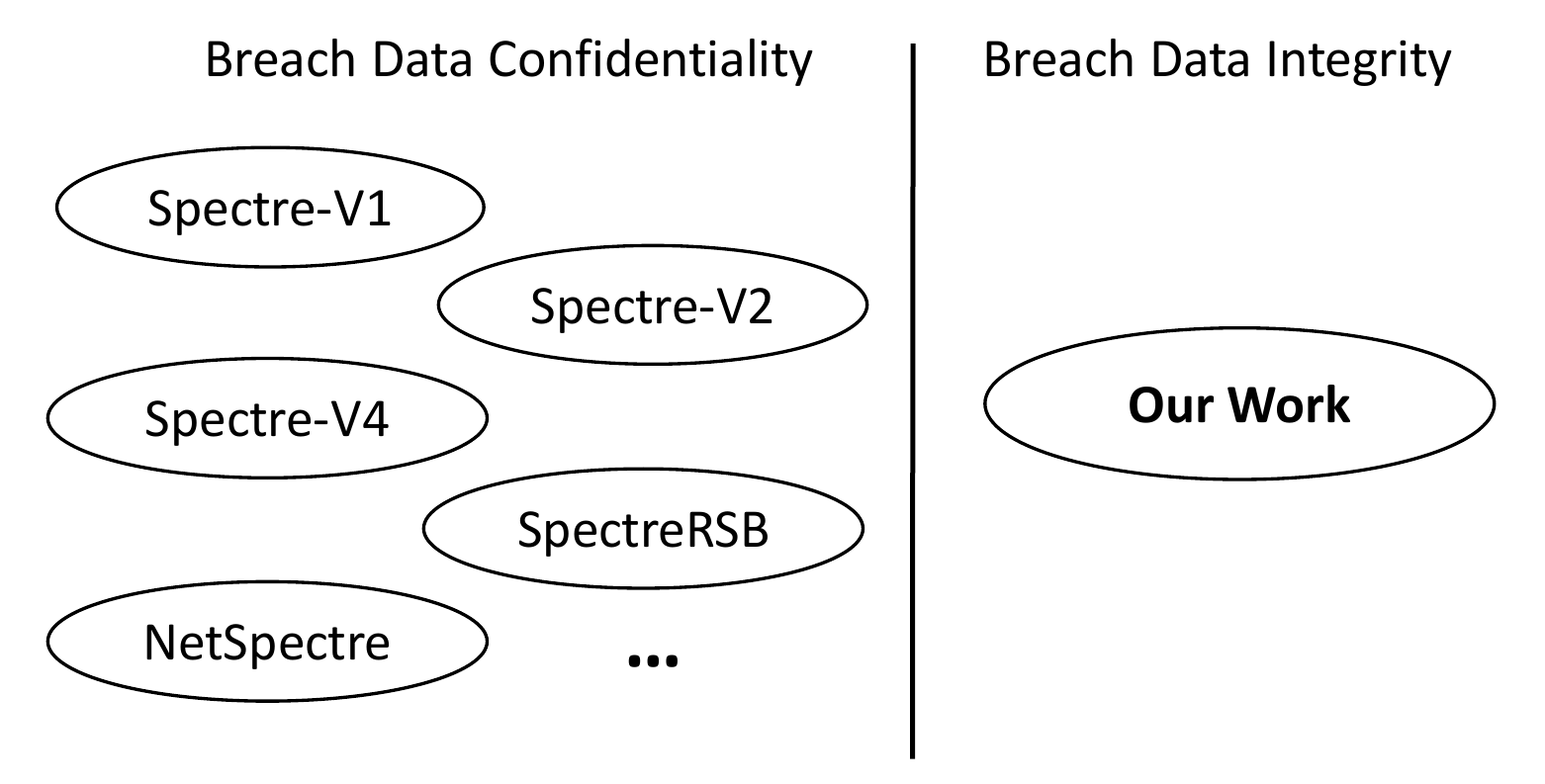}
\caption{All existing speculative execution attacks (shown on left side of the vertical line) are limited to breaching data confidentiality, also known as spectre-type attacks. In contrast to them, our work, \name, on the right side, can breach data integrity.} 
\label{fig:specpwn}
\end{figure}

Motivated by the spectre, numerous spectre-type attacks have been disclosed (e.g., spectre V1, V2, V4~\cite{kocher2018spectre,spectrev2,spectrev4}). As shown in Figure~\ref{fig:specpwn}, all spectre-like attacks in common demonstrate how to read out sensitive bits belonging to other processes, that is, the \emph{only} capability of existing speculative execution attacks is limited to breaking data confidentiality of other domains (e.g., kernel, hypervisor and etc.). 

%In the case of currently known SEAs, the attacker’s aim is limited to breaching confidentiality of data residing beyond the privilege boundary by either accessing arbitrary data or leaking specific metadata, such as pointer values, of the running program

\mypara{Our contributions}
to the best of our knowledge, we are the first to exploit the speculative execution to breach data integrity (i.e., flip bits in sensitive memory) as shown on the right side of Figure~\ref{fig:specpwn}].
We observe that the speculative execution can be triggered to leave side effects of illegal DRAM accesses, which makes it possible to break data integrity. 
%an attacker can cross memory boundary and access a target address in DRAM when  occurs.
Take the aforementioned conditional branch as an example, it checks if an array index is within the array size. If the index value passes the check, then we take the branch and perform a valid array access. 
When the array size resides within DRAM and takes about a few hundred clock cycles to arrive, the processor will not wait for the check to be resolved. Instead, it relies on prediction history records to predicts whether the check will pass. By training the branch predictor, attackers can induce the processor to speculatively perform a memory access with an illegal index that is out-of-bound. In this case, the access searches CPU cache first and then triggers a DRAM access if cache miss occurs. 
If we perform the speculative DRAM accesses as frequent as possible, then we can trigger a software-induced hardware bug~\cite{kim2014flipping} and flip bits in inaccessible.

With the above key observation, we introduce a paradigm shift in the speculative execution attacks, called \name. It is able to cross privilege boundary and flip bits in memory belonging to other domains, thus extending the capability to compromising data integrity. 
Similar to spectre-based attacks, \name also abuses the speculative execution to break existing privilege boundaries enforced by such as Memory Management Unit (MMU), MMU virtualization (e.g., Intel EPT~\cite{intelvt}, AMD NPT~\cite{amdvt}, and ARM Second-stage translation~\cite{armvt}), Intel MPK~\cite{intelvt}, Intel SGX~\cite{anati2013innovative}, AMD SME~\cite{kaplan2016amd} and etc. 

%Besides, \name can defeat practical software-only rowhammer defenses (e.g., ~\cite{brasser17can, konoth2018zebram, wu2018CAT, bock2019rip}).Since hardware mitigation such as target row refresh (TRR)~\cite{lpddr4} require DRAM upgrade and cannot be backported, software-only defenses above can protect legacy systems. They rely on MMU to enforce domain boundary and separate DRAM rows of different domains from each other so that rowhammer cannot flip bits in other domains, defending rowhammer attacks that require an attacker's DRAM rows adjacent to a target domain's rows. Nevertheless, \name can bypass the domain boundary and hammer the target domain speculatively without having DRAM adjacency between itself and the target domain. 
%leaving any architectural traces.  

%By analyzing the key requirements that consist of rowhammer~\cite{kim2014flipping} and spectre~\cite{kocher2018spectre}, respectively, 
In order to demonstrate the general idea of \name, we need to address the three following challenges.
%that need to be addressed. 
%and proposed corresponding solutions as well.
%1. trigger speculative execution \emph{reliably}.
%2. speculative window is large enough to allow memory fetch in DRAM.
%mis-speculation window: The time window during which the CPU speculatively executes the wrong code and has not yet detected that mis-speculation has occurred.
%3. the fetch is performed at a high frequency.

%\name must \emph{efficiently} and \emph{effectively} perform speculative memory accesses and 
%Recently, Intel and AMD have both proposed hard-ware supports for memory encryption within their manufac-tured processors.  Intel proposed the Software Guard Ex-tensions (SGX) [1], to protect pieces of application logic in-side encrypted enclave memory against malicious OS. How-ever SGX is limited to protect a relatively small portionof memory, and the developers have to mostly reconstructthe protected software or build it from scratch. Thus it isnontrivial for SGX to protect large-scale software like theoperating system or even the entire virtual machine. AMDproposed another simpler mechanism called Secure MemoryEncryption (SME), with which enabled, the memory can beencrypted in page level granularity by simply setting the C-bit in the page table entry

\emph{First}, \name must efficiently mistrain the processor's prediction logic in order to trigger the speculative execution as quickly as possible, since we require a high enough DRAM-access frequency to induce bit flips~\cite{kim2014flipping}. To address that, \name leverages spectre-V1 as a case and mistrains the branch predictor with a minimal number of valid array indexes. We can use one Intel Performance Counter (PMC) to decide the minimal array index number that ensures the effectiveness in mistraining CPU's branch predictor. 

%\emph{Last}, speculative DRAM access is required to be as frequent as possible within each speculation window,  In order to achieve the frequency, we minimize the effort for the nest removal as well as the aforementioned mistraining operations. 
\emph{Second}, \name must verify whether a triggered speculative execution indeed performs a DRAM access beyond memory boundary. 
%In the window, the processor is mistrained to speculatively cross memory boundary and access an invalid address from DRAM. 
If the speculation window is restricted, then the DRAM access will not occur in the window. As such, we propose a working Algorithm~\ref{alg:profile} to perform the verification. The algorithm is based on the timing difference between cache and DRAM access. 
The experimental results indicate that existing spectre-V1 attacks cannot reliably trigger speculative DRAM access, indicating that the speculation window must be extended. 

\emph{Third}, \name must reliably and efficiently extend speculation window to trigger speculative execution-based bit flips. A major reason why our triggered speculation window length is restricted is that the window is nested by preceding speculation windows. To this end, we apply previous works~\cite{mambretti2019speculator,intelOp} to terminate the preceding windows right before triggering our speculative DRAM access. However, such works cannot work in our experimental setting. Instead, 
we use an additional but empty loop to reliably stop the preceding windows. The loop count is minimized to maximize the efficiency of speculative DRAM access.

We implement \name on Ubuntu Linux and it is able to trigger the aforementioned bug and make architectural
changes, i.e., permanently flip bits in DRAM. 
%Essentially, We can develop \name-based attacks to compromise 
In our future work, we target a 1024-bit RSA exponentiation implementation running in a trusted execution environment (TEE). The TEE is provided by Intel EPT and prevents accesses from malicious users and even kernel. We leverage \name to cross the TEE, write exploitable bits into secret exponent of the RSA algorithm and gain a controllable signature.

The main contributions of this paper are as follows:
\begin{itemize}
    \item All currently known spectre-like attacks are limited to breaching data confidentiality. In contrast to them, \name undermines data integrity through speculative execution. 
    \item We present \name on Ubuntu Linux and induce permanent bit flips in DRAM, making it possible to write exploitable bits in memory of other domains. 
    %allows an unprivileged attacker to write exploitable bits into a TEE that is strictly protected by Intel EPT.
    %\item As a case, we demonstrate a \name-based attack that allows an unprivileged attacker to write exploitable bits into a TEE that is strictly protected by Intel EPT.
\end{itemize}

The rest of the paper is structured as follows.
In Section~\ref{sec:bkgd}, we briefly introduce the background information. 
In Section~\ref{sec:overview}, we present \name in detail.
%Section~\ref{sec:impl} demonstrates the exploit and evaluates it. 
%In Section~\ref{sec:mitigation}, we propose possible improvements to the physical kernel isolation against our exploit
Section~\ref{sec:dis} discusses how to utilize \name to cross different memory boundaries. Particularly, we will compromise a trusted execution environment enforced by MMU virtualization as our future work.
Section~\ref{sec:related} and Section~\ref{sec:conclusion} we summarize related works and conclude this paper, respectively.

\section{Background}\label{sec:bkgd}
\eat{
先验知识放在overview，包括哪些部分的知识。
discussion latency测量部分，比已有算法要准，
interleaved mode 有些column bits和bank bits
supports DDR3 and DDR4
limitation and advantage
a hybrid solution 
decode-dimm system knowledge 
}

In this section, we first introduce the spectre vulnerability
and then describe modern DRAM organization as well as the software-induced hardware bug.
%exploited to compromise systems without software vulnerabilities.  as well as spectre attacks
%as it is critical to understand the rowhammer bug. We then 
%summarize the existing techniques that reverse-engineer the DRAM address mapping.
 
\subsection{Spectre}
Spectre~\cite{kocher2018spectre} is a hardware vulnerability allowing an unprivileged attacker to cross memory boundary and read any target secrets. Specifically, this vulnerability resides in a Branch Prediction Unit (BPU) of most modern CPUs. The BPU enables the CPU to predict the branch target, and speculatively execute a certain number of instructions on a predicted path, the so-called ``speculative execution'' feature. 

This feature improves performance greatly particularly when the branch target is dependent on a value that is not in CPU cache but stays in DRAM. As DRAM access is much slower (a few hundred clock cycles) compared to CPU access (several clock cycles), the CPU would not idle and wait for the value to come from DRAM during this time period. Instead, it saves a checkpoint of its current valid execution state, attempts to guess the branch target based on a history of branch executions and speculatively execute instructions along the guessed path long before the value is known.
Clearly, if the educated guess turns out be wrong, then the CPU must discard architectural changes (e.g., register values) caused by the incorrectly executed instructions and revert the execution state back to the saved checkpoint for the sake of security. Unfortunately, microarchitectural side effects (e.g., CPU cache state) during the speculative execution are irrevocable, building a spectre-based timing side channel by which an adversary is able to steal a protected secret from other privileged domains (e.g., kernel, hypervisor). 

\mypara{Spectre-V1}
given that there are several spectre variants (e.g, Bounds Check Bypass and Branch Target Injection), in this paper, we utilize the variant of bounds check bypass as a case to implement \name. 
Specifically, spectre-V1 is the bounds check bypass side channel. The bounds check has conditional branch instructions used to check if an array-index candidate is within a valid range. The spectre-V1 abuses speculative execution to bypass the bounds check and speculatively access invalid memory with an out-of-bound index. The access will load invalid data into cache, which can be leaked to attackers by previous side channel techniques such as \texttt{Flush+Reload}~\cite{yarom2014flush+}.

\subsection{Dynamic Random-Access Memory}
Main memory of most modern computers uses Dynamic Random-Access Memory (DRAM). Memory modules are usually produced in the form of dual inline memory module, or DIMM, where both sides of the memory module have separate electrical contacts for memory chips. Each memory module is directly connected to the CPU's memory controller through one of the two channels. Logically, each memory module consists of two ranks, corresponding to its two sides, and each rank consists of multiple banks. A bank is structured as arrays of cells with rows and columns. Every cell stores a binary data whose value depends on whether the cell is electrically charged or not.

The charge in the DRAM cell is not persistent and will drain over time due to various charge leakage reasons~\cite{kim2014flipping}. To prevent data loss, a periodic re-charge or refresh is required for all cells. DRAM specification specifies that the DRAM refresh interval is typically 32 or 64 \emph{ms}, during which all cells within a rank will be refreshed. The higher interval indicates a better performance and thus 64\emph{ms} is the default one.

Whenever a memory access to a desired bank occurs, this ``opens'' a specified row by transferring all data in the row to the bank's row buffer and a specified column from the row buffer will be accessed. As such, subsequent to the same row will be served by the row buffer, while accessing/opening another row will flush the row buffer. 

%\mypara{DRAM Refresh}
 
%a DDR3 DRAM rank, the memory controller issues 8192 refresh commands during 64 \emph{ms}, once every 7.8us (=64ms/8192) [34].

%\mypara{DRAM Address Mapping}
%The CPU's memory controller decides how a physical address is mapped to a DRAM address. As a memory access is uniquely served by the intersection of a row and a column, a DRAM address refers to a 6-tuple of \emph{{channel}, {DIMM}, {rank}, {bank}, {row}, {column}}. However, this mapping is not publicly documented on major processor platforms such as Intel and ARM. As such, Seaborn et al.~\cite{seaborndram} observed that only different rows within the same bank can induce bit flips, based on which they made an educated guess on the DRAM address mapping of an Intel Sandy Bridge CPU. Taking a step further, both Pessl et al.~\cite{pessl2016drama} and Xiao et al.~\cite{xiao2016one} relied on a timing channel~~\cite{moscibroda2007memory} and reverse-engineered the address mapping. Pessl et al. revealed the DRAM address mapping while Xiao et al. only uncovered the mapping between a given physical address and a partial DRAM address (i.e., a 3-tuple of \emph{{bank}, {row}, {column}}).

%Google project zero
%rowhammerjs
%xiao et al.
%dramma

%\subsection{Rowhammer Overview}
\subsection{A Software-induced Hardware Bug}
Kim et al.~\cite{kim2014flipping} report the hardware bug that DRAM rows are vulnerable to persistent charge leakage induced by adjacent rows. They leverage FPGA to frequently open (also known as rowhammer) one row within the DRAM refresh interval, resulting in bit flips in a neighboring row. To trigger the bug from modern processors, memory accesses initiated by processors must also frequently reach a targeted row. As such, an adversary has to clear the CPU caches and the row buffer, and gain knowledge of how DRAM is accessed by the CPU. 

\emph{Firstly}, modern CPUs have multiple levels of caches to effectively reduce the memory access time. If data is present in the CPU cache, accessing it will be fulfilled by the cache and never reach the DRAM memory. To this end, CPU cache must be flushed in order to hammer rows and this can be done explicitly by an unprivileged instruction (e.g., \texttt{clflush}) on x86 or implicitly by eviction sets of physical pages~\cite{aweke2016anvil, rowhammerjs, bosman2016dedup,liu2015last,maurice2017hello}.

\emph{Secondly}, since the row buffer facilitates the memory accesses to the same row, bypassing it is also a necessity for hammering. As mentioned above, hammering two different rows within the same bank in an alternate manner can bypass the row buffer. If the two rows happen to be one row apart, such technique is called \emph{double-sided} rowhammer. If not, then it is coined \emph{single-sided} rowhammer. Alternatively, \emph{one-location} rowhammer~\cite{gruss2017another} forces the memory controller to clear the row buffer and thus only needs to hammer one row.

\emph{Lastly}, as all mainstream operating systems implement memory isolation, virtual addresses are the way that almost all programs running on the CPU access memory. To map a virtual address to a DRAM address, CPU's Memory Management Unit (MMU) will translate the virtual address to a physical address, which the memory controller will then map to a DRAM address. The virtual to physical mapping can be addressed by accessing \texttt{pagemap} or forcing huge-page allocation. Note that unprivileged users can access the \texttt{pagemap} interface before Linux kernel 4.0~\cite{shutemovpagemap} and they cannot allocate huge pages since the \texttt{superpage} feature is disabled by default. 

%As a result, a precise and complete DRAM address mapping efficiently induces more rowhammer bit flips whereas a higher number of bit flips increases the efficiency and effectiveness for the rowhammer attacks~\cite{gruss2017another,seaborn2015exploiting, aweke2016anvil, van2016drammer,razavi2016flip,xiao2016one,frigo2018grand,tatar2018throwhammer,lipp2018nethammer,cheng2019cattmew}.

%\subsection{Rowhammer Exploits}Published Rowhammer exploits~\cite{} go through three phases. They first hammer and scan memory for exploitable bit flips;  each memory page stores many thousands of bits, of which only a few are useful to  the  attack  in  anyway if flipped. If a bit flip , Our work focuses on providing a robust revers-engineering tool that help its users understand the rowhammer vulnerability better. 

\eat{
\subsection{Assumption}
%spectre cross boundary transient execution cross boundary works. a specific memory address range is isolated isolated environment mprotect effectiveness. kernel, MPK, hypervisor, SGX enclave, SME the Secure Encrypted Virtualization (SEV) feature, 
In order to verify whether spectre-like attacks indeed induce cross-memory-boundary rowhammer bit flips, we have the following assumptions. After confirming that \name indeed works, we discuss how to remove the assumptions. 
\begin{itemize}
    \item \item The attacker has access permission to  {\tt pagemap}. That is, the attacker can obtain the virtual-to-physical address mapping. 
\end{itemize}
}

\section{Overview}\label{sec:overview}
Our primary goal is to present the general idea of \name and demonstrate it in a real-world system. 
In this section, we firstly present the threat model and assumptions, then identify the main challenges of \name and introduce new techniques to overcome the challenges.

\subsection{Threat Model and Assumptions}
\begin{itemize}
    \item The attacker controls an unprivileged user process that has no special privileges such as accessing  {\tt pagemap} or enabling {\tt superpage}. That is, the attacker cannot obtain the virtual-to-physical address mapping. 
    \item The installed memory modules are susceptible to the software-induced hardware bug~\cite{kim2014flipping}. Pessl et al.~\cite{pessl2016drama} report that many mainstream DRAM manufacturers have vulnerable DRAM modules,  including both DDR3 and DDR4 memory.
\end{itemize}

%on each address and normal hammer on the other (i.e., repeatedly \texttt{clflush} the address and access it) to trigger the rowhammer bit flips. spectre Variant 1 (V1)~\cite{kocher2018spectre} mistrains the branch predictor with enough valid array indexes while 
\subsection{Main Challenges}
In general, we have three following steps about \name.
\circled{1} collect enough vulnerable memory addresses;
\circled{2} perform speculative hammering for each pair of addresses; 
\circled{3} check if any bit flip occurs. If not, go to step \circled{2}.
In the first step, we scan the available system memory and conduct double-sided rowhammer tool~\footnote{
https://github.com/google/rowhammer-test} to collect many enough pairs of vulnerable memory addresses that trigger bit flips. In the second step, each speculative hammering includes effectively and efficiently mistraining the CPU's prediction logic and triggering one speculative DRAM access. 
In the following, we talk about the challenges of the speculative hammering.

\mypara{efficient mistraining}
previous spectre-type attacks effectively mistrain the 
CPU's prediction logic and how to perform effective mistraining vary among spectre variants. For instance, spectre-V2~\cite{kocher2018spectre} feeds the branch predictor with enough malicious destinations.
\name can leverage any spectre variant and take spectre-V1 as an example. To this end, \name effectively mistrains the conditional branch logic of spectre-V1 with enough legitimate array indexes. 

As triggering the hardware bug requires high-frequency of DRAM accesses, the mistraining efficiency should be maximized. We empirically determine the minimum number of array indexes by using a specific Intel Performance Counter (PMC). The Intel PMC counts the mispredicted conditional branch event~\cite{intelOp} (i.e., \texttt{BR\_MISP\_EXEC.TAKEN\_CONDITIONAL}). Specifically, we first develop a kernel module to record the event count reported by the spectre-V1 proof-of-concept code (PoC)~\cite{kocher2018spectre} as the baseline, and then reduce the valid array index number one by one from the PoC until the event count is below the baseline. 
We conduct such experiment on Lenovo Thinkpad T420 with 2.6GHz Intel Core i5 2540M and 8GB DDR3 memory, and the results show that the minimal number that ensures effective mistraining can be reduced from 5 to 4. 
%In our experiments, 
%so that its speculation occurrence can be monitored by the 
%develop a kernel module that relies on the PMC to count the events each time the number is reduced by one.

%Specifically, when the number of array indexes is below a threshold, no ) can be observed by its corresponding PMC. As such, 
\begin{algorithm}[t]
\small
	\caption{Verify a speculative DRAM access}\label{alg:profile}
	\textbf{Initially:} $vul\_addr$ is a vulnerable virtual address that needs speculative hammering. $victim\_array$ is a pre-allocated array and its size is $array\_size$. $threshold$ is a predefined access latency, an indicator of a cache or DRAM access. \\
	\SetKwProg{Fn}{Function}{}{}
	    \Fn {$victim\_function (index)$} {
	        \If {$index < array\_size$} {
	            access $victim\_array + index$
	        }
	   }
	   flush both $array\_size$ and $vul\_addr$. \\
	   invoke $victim\_function$ with 4 valid indexes in sequence. \\ 
	   \Comment{efficient and effective mistraining.} \\
	   invoke $victim\_function$ with an invalid index. \\
	   \Comment{trigger speculative execution that will access $vul\_addr$.} \\
	   $latency \leftarrow$ profiling access to $vul\_addr$ \\
	   \If {$latency < threshold$} {
	    \KwRet $1$ 
	    \Comment{speculative DRAM access succeeds.}
	   }
	   \KwRet $0$ 
	   \Comment{speculative DRAM access fails.}
\end{algorithm}

\mypara{verifiable speculative DRAM access}
after the mistraining with a minimal number of valid array indexes, we can trigger the speculative execution with an invalid array index, which points to one vulnerable address. As the invalid array index needs \texttt{clflush} and resides within DRAM, there exists an speculation window between when the conditional branch instruction is issued and when it is committed. With the window, the processor is expected to speculatively access the vulnerable address from DRAM. To ensure that the access occurs in DRAM rather than cache, we \texttt{clflush} the vulnerable address before each speculative window. Note that if the window is narrow, then the processor cannot complete the DRAM access and thus fails to trigger the hardware bug. As the aforementioned Intel PMC is only able to indicate whether the speculation window exists, we cannot know whether the window is large enough. To this end, we propose Algorithm~\ref{alg:profile} to verify whether a speculative DRAM access occurs within the window.

Typically, modern processors have multiple levels of cache and external physical memory (i.e., DRAM) to manage memory accesses. 
Compared to DRAM, caches are much smaller but faster memory that store copies of frequently-accessed values. To be specific, loading a value from caches costs no more than 100 cycles while fetching it from DRAM often takes several hundred cycles. 
When a value is fetched from DRAM, a copy of that value and its nearby values in DRAM will be placed into the cache, to reduce the latency of future accesses. 

As such, we mistrain the branch predictor logic inside $victim\_function$ in line 7 and then trigger the speculative execution in line 9, with the hope that $vul\_addr$ pointed by the invalid index will be accessed from DRAM once. To check whether the DRAM access occurs, we profile the access latency to the vulnerable address in line 11. If the access latency is less than a predefined $threshold$ (e.g., 100 cycles) in line 12, then $vul\_addr$ is speculatively accessed from DRAM and thus cached. Otherwise, the speculative DRAM access fails and returns 0.  

\begin{figure}
\centering
\includegraphics[width=\columnwidth]{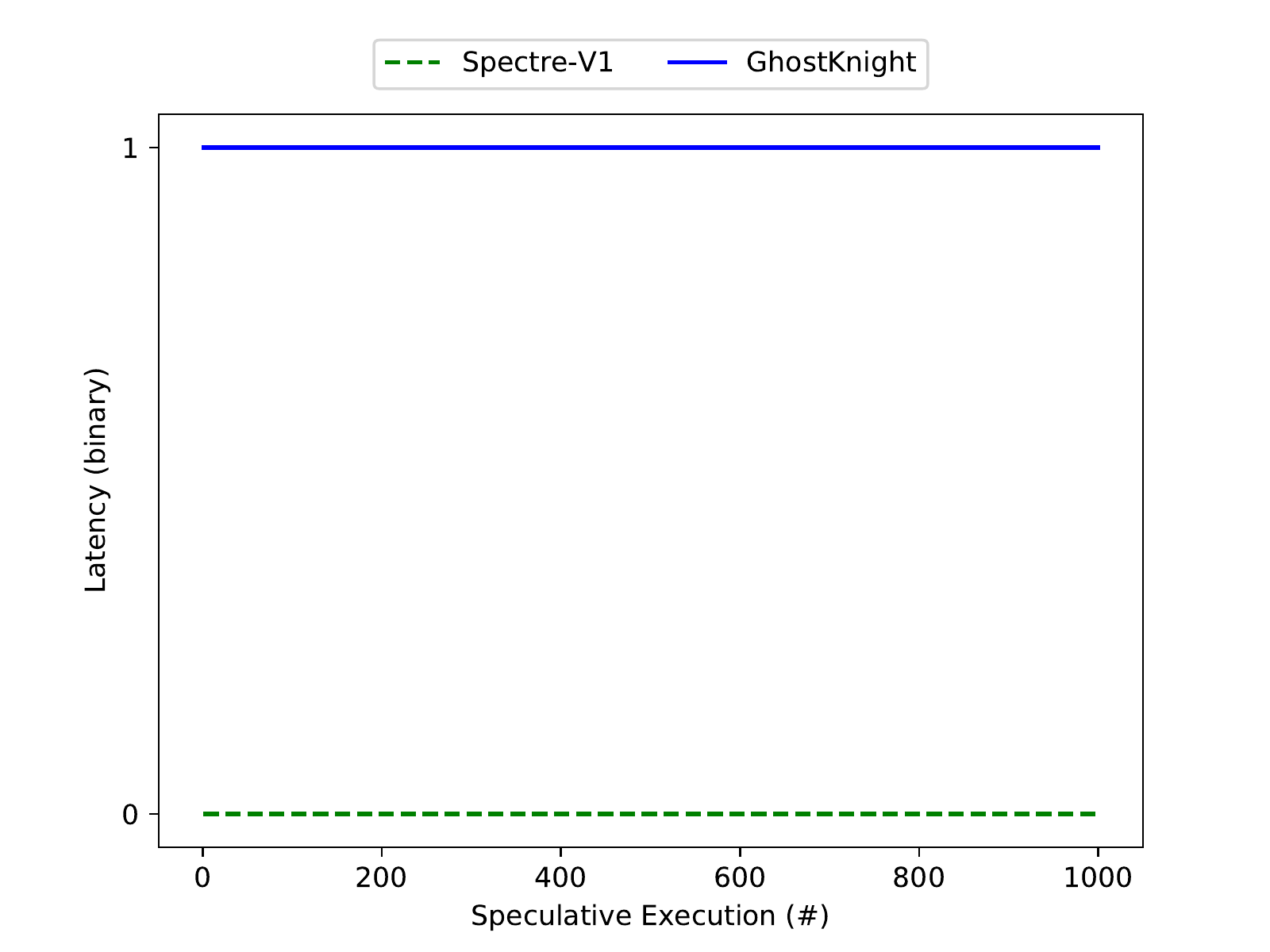}
\caption{0 in Y-axis means that a speculative DRAM access showed in X-axis does not occur. Current spectre-V1 Attack~\cite{kocher2018spectre} has a small speculation window and cannot speculatively access a target address from DRAM as described in the dashed line. \name extends the window and reliably accesses the address from DRAM each time as shown in the solid line.} 
\label{fig:wonest}
\end{figure}

We implement the Algorithm on Lenovo T420 based on the above spectre-V1 PoC~\cite{kocher2018spectre} and the results are displayed in the dashed line in Figure~\ref{fig:wonest}. Clearly, we cannot observe a single one speculative DRAM access out of 1000 speculative executions by leveraging the existing spectre-V1 attack~\cite{kocher2018spectre}. 

\mypara{reliable and efficient speculative DRAM access}
as the spectre-V1 attack cannot reliably hammer the vulnerable address, the speculation window needs to be extended. A possible reason why the window length is restricted is because the window is nested into a preceding speculation window which might be nested into another one. 

To present a long enough speculation window for one DRAM access,
previous works~\cite{mambretti2019speculator,intelOp} use instructions such as \texttt{mfence} or \texttt{lfence} to  terminate the preceding windows each time before triggering the speculative DRAM access. However, most of the 1000 speculative
executions still satisfy the address access from cache, indicating that such works do not work at least in our test machine. 

To address this problem, we place one time-consuming instruction (i.e., \texttt{syscall}) before triggering the target speculative execution and the instruction is used to exhaust the preceding window. The experimental results indicate that such placement ensures 100\% speculative DRAM accesses. However, the time cost caused by the \texttt{syscall} is high, making it impossible to trigger the hardware bug. 
To address the efficiency problem, we replace the instruction with a finite but empty loop. The loop wastes as few CPU cycles as possible in order to achieve reliable and efficient speculative hammering. 
As the solid line in Figure~\ref{fig:wonest} shows, 
\name has extended the speculation window and ensures illegitimate DRAM access inside the window.

%perform preparations to increase the success chances of speculative rowhammer. Particularly,

\begin{figure*}[ht]
	\centering
	\begin{subfigure}[t]{\columnwidth}
		\centering
		\includegraphics[scale=0.5]{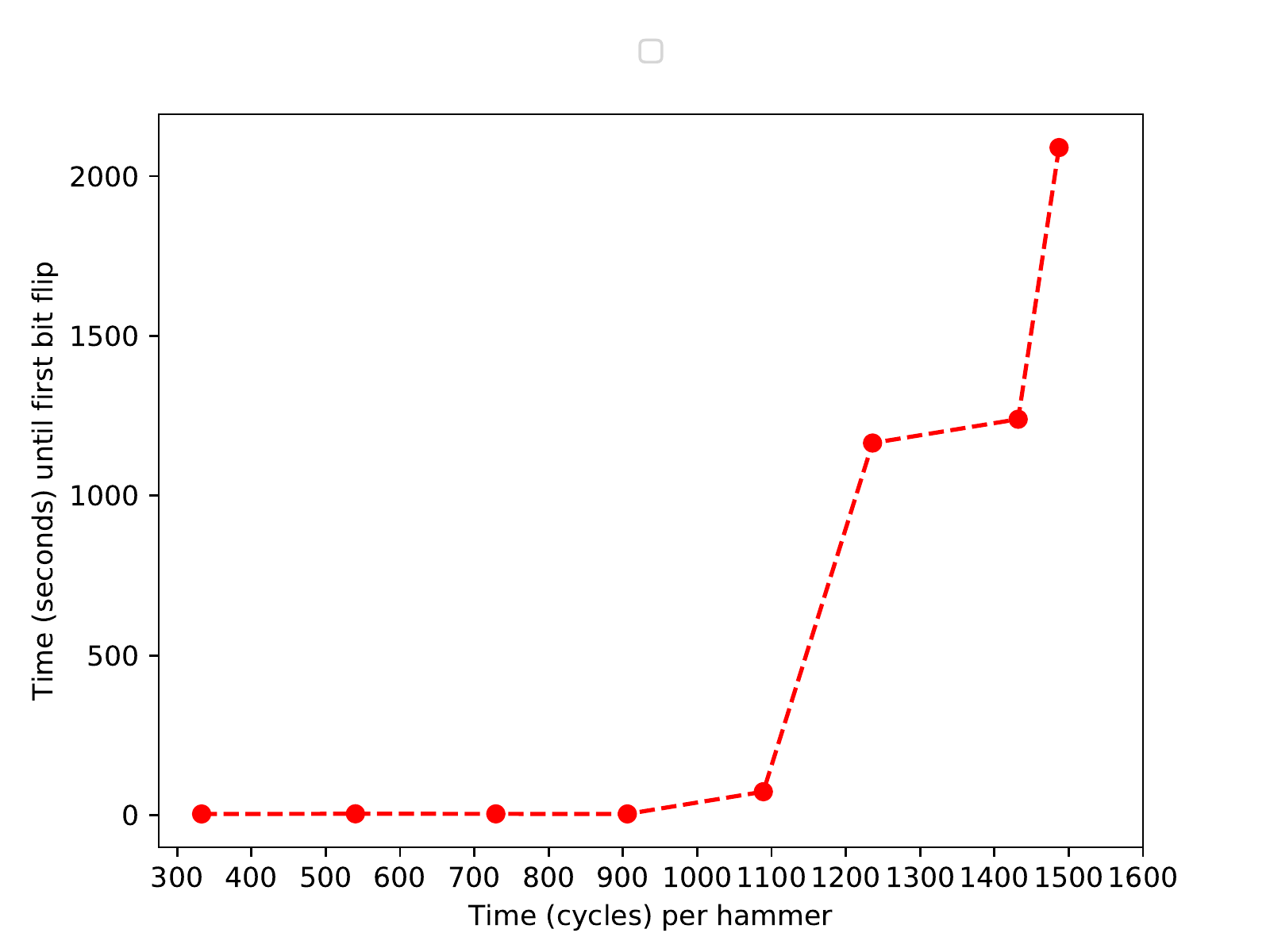}
		\caption{As the time cost per double-sided hammer increases, the time to find the first bit flip also grows. When the time cost per hammer is greater than 1500 cycles, no bit flip is observed within 2 hours.}
		\label{fig:bitflip}
	\end{subfigure}
	\hfill
	\begin{subfigure}[t]{\columnwidth}
		\centering
		\includegraphics[scale=0.5]{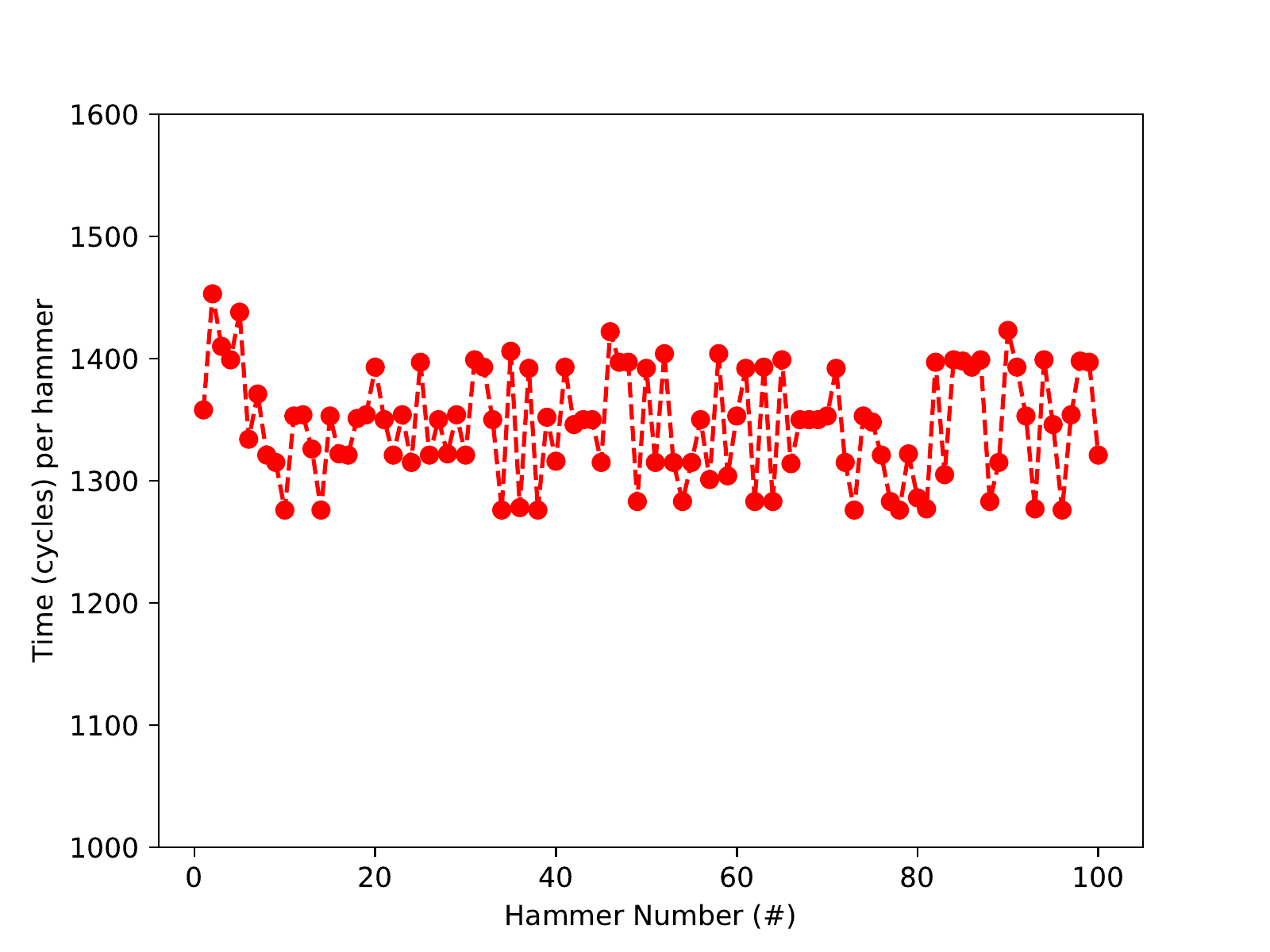}
		\caption{100\% of the time costs per speculative hammering are less than 1500. 92\% of the time costs are within the range of \{1200, 1400\}.}
		\label{fig:timecost}
	\end{subfigure}%
	\caption{The time cost caused by one speculative hammering is clearly below the maximum cost that allows bit flips, indicating that \name is efficient enough to induce bit flips.}
	\label{fig:cyclerange}
\end{figure*}

\subsection{Time cost per speculative hammering}
After addressing the above challenges, we profile the time cost taken by one speculative hammering to verify whether \name is efficient enough to trigger the hardware bug and flip bits. 
%Note that the time cost comes from both mistraining the CPU's branch predictor and triggering speculative DRAM access to the vulnerable address.  
%In the following, we are going to whether \name  
%to check we time cost for each hammer must be no greater than the maximum latency allowed to induce bit flips
Given that double-sided hammering is the most efficient way to flip bits, we use the aforementioned rowhammer tool to determine the maximum time cost that permits a bit flip on the test machine.
%through the . 

As the \texttt{clflush} instruction is the most efficient (costs below 200 cycles) and effective (cache miss rate per one round is 100\%) way to flush all levels of CPU caches, we modify the tool to embed two such instructions inside one round of double-sided hammering. In order to increase the time cost of each round, we put a certain number of \texttt{NOP} instructions preceding the \texttt{clflush} instructions per one hammering. We incrementally add the \texttt{NOP} number so that the time cost per hammering will grow. The time cost for the first bit flip to occur is shown in Figure~\ref{fig:timecost}. As we can see from the Figure, when the \texttt{NOP} number is 0, the first bit flip is observed within 10 seconds. 
As the \texttt{NOP} number grows, the time costs until the first bit flip also increase.
When the time cost per hammer is more than 1500 cycles, we cannot observe the bit flip within 2 hours. As such, 1500 can be the maximum cost that allows bit flips.

We then check whether the time taken per speculative hammering is less than the maximum cost. Unfortunately, speculative hammering one address is costly, making speculative hammering a pair of addresses hard to meet the time requirement. Instead, \name applies speculative hammering on one address and direct hammering on the other address, and the time cost of hammering the address pair is displayed in Figure~\ref{fig:cyclerange}. All the time costs per speculative hammering are well below 1500, indicating that \name is efficient to induce bit flips. 
In our experiments, \name can induce the first bit flip within 5 minutes of speculative hammering. 
Alternatively, \name can leverage the one-location hammering to speculatively hammer only one address.  

\section{Discussion}\label{sec:dis}
In this section, we first talk about our future work, that is, compromise a trusted execution environment (TEE) enforced by MMU virtualization and then we discuss how to cross other privilege boundaries. 

\subsection{Attacking MMU Virtualization}
\mypara{MMU Virtualization}
in an MMU-assisted virtualization environment, there are two levels of page tables.
The first-level page table, i.e., Guest Page Table (GPT), is managed by the kernel in the guest space, and the other one, e.g., Intel's Extended Page Table (EPT), AMD's Nested Page Table (NPT) or ARM's Second-stage Page Table, is managed by the hypervisor in the hypervisor space.
The hardware checks the access permissions at both levels for a memory access.
If the hypervisor removes the executable permission for a page $P_a$ in the EPT, then the page $P_a$ can never be executed, regardless of its access permissions in the GPT.

As a result, such hypervisor-based access control removes the potentially compromised kernel out of Trusted Computing Base (TCB) and motivates numerous security defenses~\cite{yang2008using,cheng2013appshield,hofmann2013inktag,chen2008overshadow,garfinkel2003terra}. Such works rely on the hardware-assisted hypervisor to provide a trusted execution environment (TEE) for critical code and data of a target application, thus safeguarding data confidentiality and data integrity. Among the defenses,
%many research works have leveraged the MMU virtualization feature to provide a TEE, among which
AppShield~\cite{cheng2013appshield} is a typical example and provides integrity and confidentiality of data residing in TEE. 
AppShield is a tiny hypervisor with the MMU virtualization deployed. It is able to isolate a protected application's virtual address space and block accesses to the address space unless they are authorized by the application. The application is full-fledged without any restriction such that it can request the untrusted kernel to (de)allocate DRAM memory and access the allocated memory at native speed as in a bare-metal (unprotected) setting. 

%MMU virtualization provides a trusted execution environment (TEE) for security-sensitive applications against malicious counterparts such as kernel or other applications. 
%is hardware nested paging such as Intel EPT~\cite{intelvt} and AMD NPT~\cite{amdvt} and

%'s address space such that all accesses from the kernel are blocked except those explicitly authorized by the application through system calls.  The protected application utilizes the main memory in the same fashion as in a normal (unprotected) setting.  It ac-cesses the memory with native speed, i.e. without encryption/decryption or beingintercepted, 
%These mechanisms have been widely supported by hardware processors (e.g., Intel~\cite{intelvt}, AMD~\cite{amdvt}) and commodity OSes.

Essentially, \name can defeat all the above defenses that rely on the MMU virtualization. As a case, \name will show how to cross the privilege boundary enforced by AppShield and write exploitable bits in the TEE.  
Specifically, we place a 1024-bit RSA exponentiation implementation (using square and multiply as its exponentiation) into TEE provied by AppShield, and leverage \name to bypass AppShield's security guarantees by writing bits into the secret exponent of the algorithm, resulting in an attacker-controllable signature.

\subsection{Attacking Other Privilege Boundaries}
Also, \name potentially can bypass other known privilege boundaries and flip bits in other privilege domains. For instance,
\begin{itemize}
    \item intro-process separation (i.e., sandboxed code);
    \item inter-process separation;
    \item user-kernel separation;
    \item hardware encalve separated from user or kernel-space;
    \item remote-local process separation;
\end{itemize}

%As all such privilege boundaries also require an unprivileged adversary cross privilege boundary, 
When defeating MMU virtualization, \name can simply flush cached target vulnerable address pairs by using the \texttt{clflush} instruction although they are inaccessible. However, it is challenging for \name to flush target address pairs when breaking other privilege boundaries, since \texttt{clflush} is not available in a sandboxed environment or cannot be applied to a privileged vulnerable address.
%the vulnerable address is in the privileged domain and
%Compared to the previous attack, one of the big challenges here is to hammer a privileged  that might be cached. 
Intuitively, there are two following approaches for \name to either flush or bypass cache. 

\mypara{Eviction-based} 
an attacker can evict any target address by accessing enough congruent memory addresses which are mapped to the same cache set as the target address~\cite{aweke2016anvil, rowhammerjs, bosman2016dedup,liu2015last,maurice2017hello, zhang2019telehammer}. We can use this approach to evict a cached privileged address although we do not have the access permission. To achieve a high DRAM-access frequency, the eviction should be efficient and requires a small enough eviction set congruent to the target address.  

\mypara{Uncached memory-based} 
as direct memory access (DMA) memory is uncached, past attacks (e.g., Throwhammer~\cite{tatar2018throwhammer}, Nethammer~\cite{lipp2018nethammer} Drammer~\cite{van2016drammer}) have abused DMA memory for hammering. 
    Similarly, we might find such uncached privileged memory for speculative hammering. 
    %Specifically, the address will be cached after being speculatively accessed for the first time and subsequent speculative accesses will fetch the address from cache, thus failing to trigger the rowhammer bug. As such, it is necessary to flush the address after each speculative hammering. Unfortunately, unprivileged attacker does not have the permission to flush a privileged vulnerable address. To address this challenge, we can apply 
% \end{itemize}
 
%Besides, \name can hammer two addresses that are in different rows of the same bank so as to clear the row buffer. When hammering two privileged addresses, it is difficult to verify whether the two addresses are in the same DRAM bank as well as their respective row index. 

\section{Related Work}\label{sec:related}
In this section, we first decompose existing spectre-type attacks from a high-level and then briefly introduce the microarchitectural causes of such attacks.

\subsection{Attack Phases}
All existing spectre-like attacks have demonstrated how to abuse different speculation primitives (e.g., conditional branch direction prediction) of modern processors to read arbitrary memory. In general, these attacks can be decomposed into 4 common phases:
\begin{itemize}
\item Prepare microarchitectural side channel. As speculative executions leave side effects in microarchitectural buffers such as cache, private data can be inferred in the last phase using existing timing vectors (e.g., \texttt{Prime+Probe}~\cite{liu2015last}). To this end, microarchitectural buffer states are polluted. 
\item Prepare misspeculative execution. Depending on a specific speculation primitive, the CPU is tricked into executing code with attacker-controlled arguments. The code is usually within the context of a target privileged domain such as kernel. 

\item Trigger misspeculative execution. 
There exists a time window between when permission checks in the pipeline are issued and when they are committed or retired. To fully utilize the window, the CPU will make mispredictions based on the previous phase and execute transient instructions, resulting in permanent microarchitectural state changes but transient architectural state changes. The misspeculative execution encodes secrets of other domains through microarchitectural state changes. 

\item Read Secrets via microarchitectural side channel. In this phase, secrets are reconstructed by decoding the micro-architectural state changes. This can be done by using an existing timing vector mentioned in the first phase.
\end{itemize}

\subsection{Microarchitectural Causes}
\mypara{Pattern History Table}
spectre-V1~\cite{kocher2018spectre}, NetSpectre~\cite{schwarz2019netspectre} and SGXSpectre~\cite{o2018spectre} poison Pattern History Table (PHT) to enable branch direction misprediction. 
The PHT, a component of the branch prediction unit (BPU), is a two-dimensional table of counters and each table entry is a 2-bit saturating counter. The counter stores one of two kinds of information. One is about the virtual address bits of a recently executed branch instruction, and the other is a combination of 
the branch instruction address and the outcome of the branch (i.e., branch history)~\cite{canella2019systematic,evtyushkin2018branchscope}. Based on the PHT, the CPU can predict whether a conditional branch should be taken or not.

%Kocher et al.~\cite{kocher2018spectre} first introduced Spectre-PHT, an attack that poisons the Pattern History Table (PHT) to mispredict the direction (taken or not-taken) of conditional branches. Depending on the underlying microarchitecture, the PHT is accessed based on a combination of virtual address bits of the branch instruction plus a hidden Branch History Buffer (BHB) that accumulates global behavior for the last N branches on the same physical core.

\mypara{Branch Target Buffer}
spectre-V2~\cite{kocher2018spectre} and SGXPectre~\cite{chen2019sgxpectre} poison Branch Target Buffer (BTB) to enable branch target misprediction. The BTB is also a component of the BPU and stores target virtual addresses of \emph{N} most recently executed branches. By looking up the BTB, the CPU can directly obtain the target address and speculatively fetch corresponding instructions in the next cycle.

\mypara{Return Stack Buffer}
both Koruyeh et al.~\cite{koruyeh2018spectre} and Maisuradze et al.~\cite{maisuradze2018ret2spec} poison Return Stack Buffer (RSB) to hijack the return flow during the CPU's speculative execution.
The RSB stores the \emph{N} most recent return virtual addresses, that is, the virtual addresses following the \emph{N} most recent call instructions. 
To predict the return address before executing a \texttt{ret} instruction, the CPU first pops the top most entry from the RSB to predict the return destination. 

%CPUs remember the address of the instruction followingthe corresponding call instruction. This prediction is donevia return stack buffers (RSBs)
%When encountering a ret instruction, .When the \emph{hyperthreading} feature is enabled, RSB is a microarchitectural buffer shared Note that, in case of hyperthreading,RSBs are dedicated to a logical core. The RSB size,��, variesper microarchitecture. Most commonly, RSBs are��=16entries large, and the longest reported RSB contains��=32entries in AMD’s Bulldozer architecture [11]. In this paper, weassume an RSB size of 16, unless explicitly stated otherwise,but our principles also work for smaller or larger RSBs.

\mypara{Store To Load Dependency}
spectre-V4~\cite{spectrev4} poisons Store To Load Dependency (STLD) to trick the CPU into speculatively execute a load instruction even if it is unknown wheter the instruction is overlapped with previous store instructions.
The STLD requires that a load micro-op shall not be executed before all preceding store micro-ops complete writing to the same memory location. For the sake of performance, the CPU's memory disambiguator will predict which load does not depend on any prior stores. If there is a load that requires no such dependency, then the load will speculatively read data from the L1 data cache. When the physical addresses of all prior stores are known, the prediction is verified. If the load conflicts with at least one previous store, the load and its succeeding instructions are re-executed.

%Spectre exploits a performance optimization in modern CPUs. Instead of waiting for the correct resolution of a branch, the CPU tries to predict the most likely outcome of the branch and starts transiently executing along the predicted path. Upon resolving the branch, the CPU discards the results of the transient execution if the prediction was wrong but does not revert changes in the microarchitecture. The prediction is based on events in the past, allowing an attacker to mistrain the predictor to leak data through the microarchitecture that should normally not be accessible to the attacker.

\section{Conclusion}\label{sec:conclusion}
In this paper, we demonstrated the first exploit that utilized speculative execution to break data integrity. 
In the near future, we will demonstrate a \name-based attack to defeat the MMU virtualization and write exploitable bits beyond the privilege boundary.

\section{Acknowledgements}\label{sec:acknowledgement}
We are grateful to Yinqian Zhang (a professor from Southern University of Science and Technology) for his valuable feedback on this paper.



%\bibliographystyle{IEEEtranS}

 \bibliographystyle{IEEEtranS}
\bibliography{main}

%\begin{thebibliography}{1}

%H.~Kopka and P.~W. Daly, \emph{A Guide to \LaTeX}, 3rd~ed.\hskip 1em %plus
%\end{thebibliography}

\end{document}